\documentclass[twocolumn,showpacs,prb,citeautoscript]{revtex4}
\usepackage{graphicx}
\usepackage{dcolumn}
\usepackage{bm}
\usepackage{times}

\begin{document}

\title{Multiband model for tunneling in MgB$_2$ junctions}
\author{A. Brinkman}
\author{A.A. Golubov}
\author{H. Rogalla}
\affiliation{Department of Applied Physics and MESA+ Research
Institute,\\University of Twente, 7500 AE, Enschede, The
Netherlands}
\author{O.V. Dolgov}
\author{J. Kortus}
\author{Y. Kong}
\author{O. Jepsen}
\author{O.K. Andersen}
 \affiliation{Max-Plank Institut f{\"ur} Festk{\"o}rperforschung, Heisenbergstr. 1,
D-70569, Stuttgart, Germany}

\date{\today }

\begin{abstract}
A theoretical model for quasiparticle and Josephson tunneling in
multiband superconductors is developed and applied to
MgB$_2$-based junctions. The gap functions in different bands in
MgB$_2$ are obtained from an extended Eliashberg formalism, using
the results of band structure calculations. The temperature and
angle dependencies of MgB$_2$ tunneling spectra and the Josephson
critical current are calculated. The conditions for observing one
or two gaps are given. We argue that the model may help to settle
the current debate concerning two-band superconductivity in
MgB$_2$.
\end{abstract}
\pacs{74.50.+r, 74.70.Ad, 74.80.Fp, 85.25.Cp}
\maketitle

Soon after the discovery of superconductivity in MgB$_2$ \cite{akimitsu},
first principle calculations were performed to determine the electronic structure
of this material. It was found that the Fermi surface
consists of two three-dimensional sheets, from the $\pi$ bonding and antibonding bands, and two nearly
cylindrical sheets from the two-dimensional $\sigma$ bands \cite{Kortus,An,Kong,Bohnen}. The multiband
picture has given rise to the concept that two superconducting energy gaps can coexist \cite{Liu,Shulga} in
MgB$_2$.

Two-band superconductivity is a phenomenon that has been observed
in Nb doped SrTiO$_3$ \cite{Bednorz}. Recent experimental
STM and point-contact spectroscopy
\cite{Klein,Bugoslavsky,Szabo}, high-resolution photo-emission
spectroscopy \cite{Tsuda}, Raman spectroscopy \cite{Chen},
specific heat measurements \cite{sph} and muon-spin-relaxation
studies of the magnetic penetration depth \cite{Nieder}
support the concept of a double gap in MgB$_2$ (see Ref.\
\onlinecite{Buzea} for a review of experiments). However, there is
an ambiguity in the interpretation of point-contact data
concerning the existence of two gaps
\cite{Klein,Bugoslavsky,Szabo}. Moreover, some tunneling
measurements \cite{Gonelli} show only one gap with a magnitude
smaller than the BCS value of $\Delta =1.76$ $k_B T_c$.

In order to resolve this discrepancy, we address the question,
how multiband superconductivity will manifest itself in tunneling.
We present the theoretical model for quasiparticle and Josephson
tunneling in MgB$_2$-based junctions. Using the results of
band-structure calculations, we apply an extended Eliashberg
formalism to obtain the gap functions in different bands, taking
strong coupling effects into account. Tunneling from a normal
metal (N) into MgB$_2$ is considered in an extended
Blonder-Tinkham-Klapwijk (BTK) model \cite{BTK}. The temperature
dependencies and absolute values of the $I_cR_N$ product ($I_c$
is the critical current and $R_N$ is the normal state resistance)
are calculated in MgB$_2$-based SIS tunnel junctions, where S
denotes a superconductor and I an insulator. Tunneling in the
direction of the $a$-$b $ plane, in the $c$-axis direction and
under arbitrary angle is considered. Furthermore, the Josephson
supercurrent between a single-gap superconductor and MgB$_2$ is
calculated.

According to the labeling of Liu \textit{et al.}\ \cite{Liu}, the
four Fermi surface sheets in MgB$_2$ are grouped into
quasi-two-dimensional $\sigma $ bands and three-dimensional $\pi $
bands. Hence, normal and superconducting properties of MgB$_2$ can
be described by an \textit{effective two-band model}. Within this
model, Liu \textit{et al.}~\cite{Liu} estimated the coupling
constants and energy gap ratio in the weak coupling regime. More
recently, the band decomposition of the superconducting and
transport Eliashberg functions $\alpha_{ij}^2F_{ij}\left( \omega
\right) $ (where $i$ and $j$ denote $\sigma $ or $\pi $ bands),
which describe the electron-phonon coupling in MgB$_2$ as function
of the frequency $\omega $,  was provided in
Ref.~\onlinecite{Dolgov}. This allows to perform a strong coupling
calculation of the superconducting energy gap functions $\Delta
_i\left( \omega _n\right) $ in different bands. The functions
$\Delta _i\left( \omega _n\right) $ in turn determine the
Josephson critical current in a tunnel junction between multiband
superconductors, which is given by a straightforward
generalization of the well-known result \cite{AB} to the case of
several conducting bands \cite{MGZ} as well as strong coupling.
The critical current component for tunneling from band $i$ into
$j$ is given by

\begin{equation}
I_{ij}=\frac{\pi T}{eR_{ij}}\sum_{\omega _n}\frac{\Delta _{{\cal
L}i}\left( \omega _n\right) \Delta _{{\cal R}j}\left( \omega
_n\right) }{\sqrt{\omega _n^2+\Delta _{{\cal L}i}^2\left( \omega
_n\right) }\sqrt{\omega _n^2+\Delta _{{\cal R}j}^2\left( \omega
_n\right) }},  \label{J}
\end{equation}
where ${\cal L}$ and ${\cal R}$ denote left and right
superconductors respectively, $R_{ij}^{-1}=\min \{ R_{{\cal L}ij}^{-1},R_{{\cal R}%
ij}^{-1}\} $ is the normal-state conductance of a junction for the
bands $(i,j)$ which is given by the integral over the Fermi surface $S_{{\cal L%
}i({\cal R}j)}$
\begin{equation}
\left(R_{{\cal L}({\cal R})ij } \cal A \right)^{-1} =\frac{2e^2}\hbar \int_{v_x>0}\frac{D_{ij}v_{n,{\cal L%
}i({\cal R}j)} d^2S_{{\cal L}i({\cal R}j)}}{\left( 2\pi \right) ^3v_{F,%
{\cal L}i({\cal R}j)}},  \label{R}
\end{equation}
where $\cal A$ is the junction area, $v_n$ is the projection of the Fermi velocity $v_F$ on the direction
normal to the junction plane, and $D_{ij}$ is the probability for a quasiparticle to tunnel from band $i$ in
${\cal L}$ into band $j$ in ${\cal R}$. The total critical current is the sum of the components
$I_c=\sum_{ij}I_{ij}$.

\begin{figure}
\includegraphics *[scale=0.85]{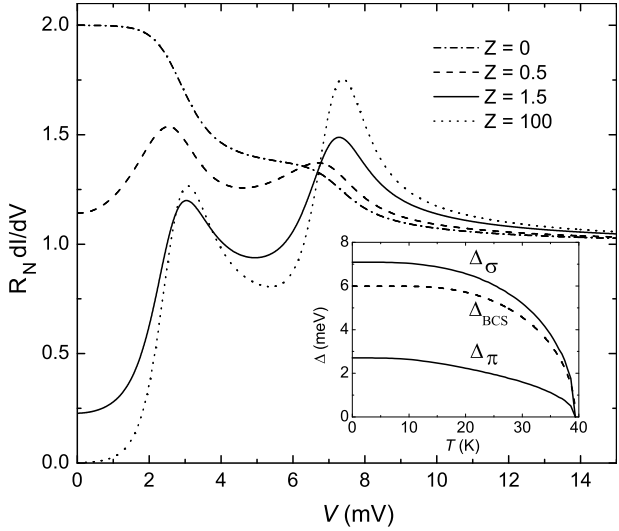}
\caption{\label{fig:fig1} The normalized conductance of
MgB$_2$-I-N junctions as function of voltage at 4.2 K, with
$Z=U_0/\hbar v_F$. The superconducting gaps for the different
bands are shown in the inset.}
\end{figure}

The gap functions $\Delta _i\left( \omega _n\right) $ can be calculated with an extension of the Eliashberg
formalism \cite{Carbotte} to two bands
\begin{equation}
\Delta _i\left( \omega _n\right) Z_i\left( \omega _n\right) =\pi
T\sum_j\sum_{\omega _m}\frac{\left( \lambda _{ij}-\widetilde{\mu}
_{ij}^{*}\right) \Delta _j\left( \omega _m\right) }{\sqrt{\omega
_m^2+\Delta _j^2\left( \omega _m\right) }},  \label{Eliashberg}
\end{equation}
\begin{eqnarray}
Z_i\left( \omega _n\right) =1+\frac{\pi T}{\omega
_n}\sum_j\sum_{\omega _m}\lambda _{ij}\frac{\omega
_m}{\sqrt{\omega _m^2+\Delta _j^2\left( \omega _m\right) }},
\label{Z}
\end{eqnarray}
where $\lambda _{ij}=2\int_0^\infty {\omega \alpha_{ij}
^2F_{ij}\left( \omega \right) d\omega }/[{\omega ^2+\left( \omega
_m-\omega _n\right) ^2}]$, $Z_i\left( \omega _n\right) $ are the
Migdal renormalization functions and $\omega _n=\pi T(2n+1)$.
These equations are solved numerically with the electron-phonon
$\alpha_{ij} ^2F_{ij}\left( \omega \right) $ functions from
Ref.~\onlinecite{Dolgov}. The cutoff frequency $\omega _c$ is
taken equal to 10 times the maximum phonon frequency. The
functions $\widetilde{\mu}_{ij}^{*}$ represent a matrix of the
Coulomb pseudopotentials defined at $\omega _c $, calculated up to
a common prefactor that is used as an adjustable parameter to get
$T_c$ = 39.4 K. The matrix $\mu^{*}$ at the frequency
$\omega_{\ln}$ (relevant for the McMillan expression for $T_c$ in
the isotropic case) is given by $\mu^{*}=[1+\widetilde{\mu}^{*}
\ln(\omega_{c}/\omega_{\ln})]^{-1} \widetilde{\mu}^{*}$, where
$\omega_{\ln}$ follows from $0=\int_0^\infty{\ln (\omega
/\omega_{\ln}) \omega^{-1} \alpha_{ij} ^2F_{ij}(\omega) d
\omega}$. The corresponding matrix elements are $\mu_{\sigma
\sigma }^{*}=0.13$, $\mu_{\sigma \pi }^{*}=0.042$, $\mu_{\pi
\sigma }^{*}=0.03$, $\mu_{\pi \pi }^{*}=0.11$ and $\lambda
_{ij}(\omega_m=\omega_n)$ from Ref.~\onlinecite{Dolgov} are $\lambda _{\sigma
\sigma }=1.017$, $\lambda _{\sigma \pi }=0.213$, $\lambda _{\pi
\sigma }=0.155$, $\lambda _{\pi \pi }=0.448$. Due to the interband
coupling terms in Eqs.~(\ref{Eliashberg}, \ref{Z}) both gaps close
at the same $T_c$. The resulting temperature dependencies of the
energy gaps,
$\Delta _i\left(T\right) ,$ are plotted in the inset of Fig.~\ref{fig:fig1}
and it is found that $\Delta _\sigma \left( T=0\right) $ = 7.09 meV and
$\Delta _\pi \left( T=0\right) $ = 2.70 meV, with the $2\Delta /T_c$ ratios
being equal to 4.18 and 1.59, respectively. For comparison, also
the BCS curve is shown for $T_c$ = 39.4 K. The BCS value for the
gap that corresponds to $T_c$ = 39.4 K is 6.0 meV at 0 K. It can
be seen that the temperature dependencies are qualitatively
different from the BCS temperature dependence. The ratio of the
gaps $\Delta _\sigma /\Delta _\pi $ increases for increasing
temperatures, as was experimentally observed for example in
Ref.~\onlinecite{Klein}.

The influence of impurities can be incorporated into the model. Intraband
scattering does not change the two gaps (Anderson's theorem), while the
interband scattering can be included by terms $\gamma _{ij}\Delta _j/\sqrt{%
\omega _n^2+\Delta _j^2}$, $\gamma _{ij}\omega _n/\sqrt{\omega _n^2+\Delta _j^2}$ in the Eliashberg
equations (\ref{Eliashberg}, \ref{Z}) respectively.
The smallness of $\gamma _{ij}$\ compared to $\pi T_c$
indicates that the double-gap feature should experimentally be observable,
also in thin films, even for a certain amount of impurity scattering.
A large amount of impurity scattering ($\gamma _{ij}$ exceeding the maximum phonon
frequency) will cause the gaps to converge to the same value.
From Eqs.~(\ref{Eliashberg}, \ref{Z}) and including the
scattering terms an asymptotic value of $\Delta _\sigma =\Delta _\pi =$\ 4.1 meV
and $T_c=$25.4 K is found, giving a $ 2\Delta /k_B T_c$ ratio of 3.7.

In order to obtain the normal state resistance, we have to evaluate the effective junction transparency
components $D_{ij}$. In the case of a specular barrier, $U(x)=U_0\delta (x-x_0)$, $D_{ij}$ is given by
\begin{equation}
D_{ij}=\frac{v_{n,{\cal L}i}v_{n,{\cal R}j}}{\frac 14\left( v_{n,{\cal L}%
i}+v_{n,{\cal R}j}\right) ^2+U_0^2/\hbar^2}.  \label{D}
\end{equation}
It follows from Eqs.~(\ref{R}, \ref{D}) and as first pointed
out in Ref.~\onlinecite{Mazin}, that the normal state conductance
$R_{ij}^{-1}$ in the large $ U_0$ limit is proportional to the
Fermi-surface average $\left\langle Nv^2\right\rangle _i$. The
latter
is proportional to the contribution of
the electrons in band $i$ to the squared plasma frequency $ \left(
\omega _p^i\right) ^2$. This essentially simplifies the task of
summing up the interband currents since the partial plasma
frequencies are available from the band structure calculations
\cite{Kortus,An,Kong,Bohnen}. The normal state junction
conductance is thus proportional to $\left( \omega _p^i\right)
_{{\cal L}}^2\left\langle v_n\right\rangle _{{\cal R}j}${\em,}
where $\left\langle v_n\right\rangle _{{\cal R}j}$ is the average
Fermi velocity projection in the corresponding band (see
Table~\ref{tab:table1}). In order to sum up the contributions of
different bands, we restrict ourselves to the weighing
factors $\left( \omega _p^i\right) ^2$, neglecting the difference
in $\left\langle v_n\right\rangle _{{\cal R}j}$. This is a
reasonable approximation since the difference between $v_F$ in the
$\sigma $ and $\pi $ bands in the $a$-$b$ plane is rather small,
while for $c$-axis tunneling only the $\pi $ band contributes, as
will be shown later, so that the problem of summation does not
appear in this case.

\begin{table}
\caption{Calculated plasma frequencies, average Fermi velocities and gap values for the $\sigma $ and $\pi $
bands. }
\label{tab:table1}%
\begin{ruledtabular}
\begin{tabular}{cccccc}
&$\omega _p^{a-b}$ (eV)&$\omega _p^c$ (eV)&$v_F^{a-b}$(m/s)&$v_F^{c}$(m/s)&$\Delta$ (meV)\\
\hline
$\sigma $ band &4.14&0.68&4.40 $10^5$&0.72 $10^5$&7.09\\
$\pi $ band &5.89&6.85&5.35 $10^5$&6.23 $10^5$&2.70\\
\end{tabular}
\end{ruledtabular}
\end{table}

\textbf{SIN tunneling.} The conductance in a MgB$_2$-I-N tunnel junction is the sum of the contributions of
two bands. Each of the conductances is given by the BTK model \cite{BTK}, where the corresponding normal
state conductances $R_{Ni}^{-1}$ are proportional to the minimum of the square of the plasma frequencies at
the N and MgB$_2$ sides. Since the plasma frequency in a typical normal metal (e.g. Au, Ag) is larger than
the plasma frequencies in MgB$_2$, the conductances are limited by the electrons on the MgB$_2$ side
\begin{equation}
R_{N\sigma }^{-1}/R_{N\pi }^{-1}=\left( \omega _p^\sigma \right)
^2/\left( \omega _p^\pi \right) ^2.  \label{R2}
\end{equation}
Finally, the normalized conductance of an N-I-MgB$_2$ contact is given by
\begin{equation}
\sigma \left( V\right) \equiv \frac{\left( \frac{dI}{dV}\right) _{NIS}}{%
\left( \frac{dI}{dV}\right) _{NIN}}=\frac{(\omega _p^\pi
)^2\sigma _\pi
\left( V\right) +(\omega _p^\sigma )^2\sigma _\sigma \left( V\right) }{%
(\omega _p^\sigma )^2+(\omega _p^\pi )^2}.  \label{conductance}
\end{equation}
Here, the dimensionless conductances $\sigma _{\sigma ,\pi }\left( V\right) $ are provided by the BTK model,
with the calculated values for the gaps and plasma frequencies, as shown in Table~\ref{tab:table1}.

In the conductance versus voltage plot (Fig.~\ref{fig:fig1}) for tunneling in the
$a$-$b$ direction, two peaks are clearly visible, in qualitative agreement with
the experimental data \cite{Klein,Bugoslavsky,Szabo}.
The ratio of peak magnitudes is not only determined by the ratio of the
plasma frequencies, but also by thermal rounding and by the barrier strength
$Z_{\text{BTK}}=U_0/\hbar v_F$ (where $v_F$ is taken constant for the
different bands for the same reason as was given in the determination of
$R_{ij}$). In particular, the peak at the smaller gap dominates in the small
$Z_{\text{BTK}}$ regime (point-contact), while the second peak dominates
at large values of $Z_{\text{BTK}}$ (tunneling), as may be seen in
Fig.~\ref{fig:fig1} at 4.2 K.

Due to the smallness of $\omega _p^\sigma $ in the $c$-direction, it can be seen
from Eq.~(\ref{conductance}) that the conductance in the $c$-direction is only
determined by the $\pi $ band.
In this case, no double-peak structure is expected in the conductance spectrum.
This explains, together with the dependence on $Z_{\text{BTK}}$,
why in some experiments only one peak was observed \cite{Gonelli}
or why the second peak was weak \cite{Klein,Bugoslavsky,Szabo}.

Note, that the assumption of the ratio of the normal state conductivities
being equal to the ratio of the square of the plasma frequencies holds
when the interface is a $\delta$-function shaped tunnel barrier, with large
$Z_{\text{BTK}}$. This means that for small $Z_{\text{BTK}}$, the results
should be considered as a qualitative indication only.
In the latter case, as well as for other types of barriers,
a numerical integration of Eqs.~(\ref {R}, \ref{D}) must be performed.

\textbf{SIS Josephson tunneling.} We consider Josephson tunneling between two MgB$_2$ superconductors. With
the values for the plasma frequencies, $ \omega _p^\sigma <\omega _p^\pi $, this gives $R_{\sigma \pi }
=R_{\pi \sigma }=\max \left( R_{\sigma \sigma },R_{\pi \pi }\right) =R_{\sigma \sigma }$ and $R_{\sigma \pi
}/R_{\pi \pi } =R_{\sigma \sigma }/R_{\pi \pi }=( \omega _p^\pi /\omega _p^\sigma ) ^2>1$. The total
conductance is given by $R_N^{-1}=\sum_{ij}R_{ij}^{-1}$.

\begin{figure}
\includegraphics *[scale=0.85]{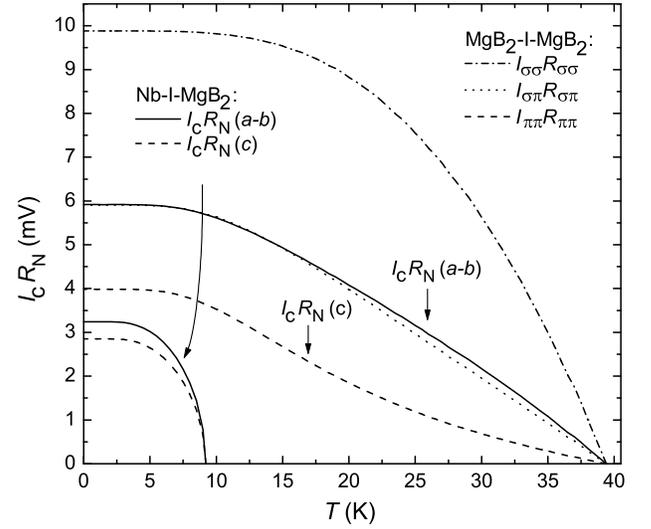}
\caption{\label{fig:fig2} $I_cR_N$ temperature dependence for
different tunneling components. The resulting $I_cR_N$ for
tunneling in the direction of the a-b plane and c-axis direction
are indicated, for MgB$_2$-I-MgB$_2$ and Nb-I-MgB$_2$ junctions
(see text).}
\end{figure}

For tunneling in the $a$-$b$ plane (as can be realized for example in an edge configuration), with
$R_{\sigma \pi } =R_{\pi \sigma }$ and $I_{\sigma \pi } =I_{\pi \sigma }$, the total $I_cR_N$ product
becomes
\begin{eqnarray}
I_cR_N =\frac{I_{\sigma \sigma }R_{\sigma \sigma }+2I_{\sigma \pi
}R_{\sigma \pi
}+I_{\pi \pi }R_{\pi \pi }\left( \omega _p^\pi /\omega _p^\sigma \right) ^2}{%
3+\left( \omega _p^\pi /\omega _p^\sigma \right) ^2}.
\label{IcRn}
\end{eqnarray}
The results of numerical calculations are presented in Fig.~\ref{fig:fig2}. Due to strong-coupling and
interband coupling effects, the temperature dependencies of $I_{ij}R_{ij}$ differ from the well-known
Ambegaokar-Baratoff result for an SIS junction between isotropic superconductors, most clearly demonstrated
by the positive curvature of the $I_{\pi \pi }R_{\pi \pi }$ contribution. The $I_cR_N$ value at $T$ = 4.2 K
is 5.9 mV.

For tunneling along the c-axis, the only contribution to the $I_cR_N$ product
comes from  $I_{\pi \pi}R_{\pi \pi }$, because of the negligible value
for $\omega _p^\sigma $ in the c-axis direction.
This gives $I_cR_N$ = 4.0 mV at $T$ = 4.2 K.

The plasma frequency in a certain direction, under an angle $\varphi $ with the $a$-$b$ plane, can be
determined from the ellipsoid equation $(\omega _p^i)^2=(\omega _{p,x}^i)^2+(\omega _{p,z}^i)^2,$ and
$(\omega_{p,z}^i/\omega _{p,c}^i)^2+(\omega _{p,x}^i/\omega _{p,a-b}^i)^2=1$, where $%
\omega _{p,x}^i$ and $\omega _{p,z}^i$ form the decomposition of $\omega _p^i $. Because of the negligible
value of $\omega _{p,c}^\sigma $, it is evident that $\omega _p^\sigma$ is negligible for nonzero values of
$\varphi =\arctan (\omega _{p,z}^i/\omega _{p,x}^i)$. This implies that tunneling under a nonzero angle with
the $a$-$b$ plane gives the same result as tunneling in the $c $-axis direction, namely $I_cR_N$ = 4.0 mV at
$T$ = 4.2 K. For angles approaching zero (of the order of 0.6$^{\circ }$), $I_cR_N$ rapidly increases
towards the maximal value for tunneling from $a$-$b$ plane to $a$-$ b$ plane, namely $I_cR_N$ = 5.9 mV at
$T$ = 4.2 K. For a large amount of impurity scattering the $I_{ij}R_{ij}$ values converge to the same value.
It follows in that case from Eq.~(\ref{IcRn}), with the plasma frequencies from Table~\ref{tab:table1}, that
$I_cR_N$ becomes almost isotropic.

Finally, tunneling from MgB$_2$ into a superconductor $S^{\prime
}$ with a single gap will be considered (we take Nb as an
example). The resulting $I_{iS^{\prime }}R_{iS^{\prime }}$
temperature dependencies are calculated numerically, using 1.4 mV
for the energy gap in Nb. The ratio of resistances is determined
from Eq.~(\ref{R}). Since typical values of plasma frequencies in
other superconductors are bigger than in MgB$_2$ (e.g. 9.47 eV for
Nb, 12.29 eV for Al and 14.93 eV for Pb, see
Ref.~\onlinecite{Maksimov}), the following expression is obtained
\begin{equation}
I_cR_N=\frac{I_{\sigma S^{\prime }}R_{\sigma S^{\prime }}+I_{\pi
S^{\prime
}}R_{\pi S^{\prime }}\left( \omega _p^\pi /\omega _p^\sigma \right) ^2}{%
1+\left( \omega _p^\pi /\omega _p^\sigma \right) ^2},  \label{Nb}
\end{equation}
when tunneling occurs into the $a$-$b$ plane of the MgB$_2$. In the case of $c$-axis tunneling, only the
$I_{\pi S^{\prime }}R_{\pi S^{\prime }}$ contribution remains. The results for tunneling from Nb to MgB$_2$
are also indicated in Fig.~\ref{fig:fig2}. Other superconductors give qualitatively similar results. The
only scaling parameter is the critical temperature of the superconducting counter-electrode.

Our results for Josephson tunneling provide an upper bound for $I_cR_N$ products,
being 5.9 mV and 4.0 mV for tunneling into the $a$-$b$ plane and $c$ direction
respectively. There have already been several
observations of Josephson currents in MgB$_2$ junctions \cite{Josephson},
with $I_cR_N$ values that are much lower than our predictions.
This can be due to extrinsic reasons such as a degradation of the $T_c$ of
surface layers in the vicinity of the barrier, the barrier nature and barrier
quality. From our model, however, it follows that polycrystallinity does
not reduce the Josephson coupling very much, as indicated by
the calculated value of $I_cR_N $ of 4.0 mV for c-axis transport,
neither does strong impurity scattering
because of the relatively large average gap of 4.1 meV in this case.

In conclusion, Josephson tunneling in MgB$_2$-based junctions is
discussed theoretically in the framework of a two-band model. The
gap functions in different electronic bands are calculated using
the Eliashberg formalism together with band structure
information. This provides a basis to interpret electronic
transport in MgB$_2$. We have shown the possibility to observe
either one or two gaps in point-contact spectra of MgB$_2$,
depending on the tunneling direction, barrier type and amount of
impurities. The results are also relevant for the electronic
application of MgB$_2$ since they provide the limit for the
Josephson coupling strength in MgB$_2$ based junctions. For
MgB$_2$ in the clean limit we have shown that $I_cR_N$ values as
high as 5.9 mV can be expected for MgB$_2$ tunnel junctions if
tunneling occurs in the direction of the $a$-$b$ plane.
In other cases the limiting $I_cR_N$ values will not exceed 4.0
mV. Our predictions for the gap and $I_cR_N$ anisotropy and for
the $I_c$ vs $T$ dependence in MgB$_2$-based junctions can be
verified experimentally and thus may help to settle the current
debate on two-band superconductivity in MgB$_2$.

The authors thank D.H.A. Blank, H. Hilgenkamp and I.I. Mazin for
useful discussions. This work was supported by the Dutch
Foundation for Research on Matter (FOM).

\end{document}